# Spatial and Spectral Coherent Control with Frequency Combs

Itan Barmes, Stefan Witte and Kjeld S. E. Eikema*

LaserLaB Amsterdam, VU University, De Boelelaan 1081, 1081HV Amsterdam, The Netherlands.

*Correspondence to: K.S.E.Eikema@vu.nl

**Quantum coherent control (1-3) is a powerful tool for steering the outcome of quantum processes towards a desired final state, by accurate manipulation of quantum interference between multiple pathways. Although coherent control techniques have found applications in many fields of science (4-9), the possibilities for spatial and high-resolution frequency control have remained limited. Here, we show that the use of counter-propagating broadband pulses enables the generation of fully controlled spatial excitation patterns. This spatial control approach also provides decoherence reduction, which allows the use of the high frequency resolution of an optical frequency comb (10,11). We exploit the counter-propagating geometry to perform spatially selective excitation of individual species in a multi-component gas mixture, as well as frequency determination of hyperfine constants of atomic rubidium with unprecedented accuracy. The combination of spectral and spatial coherent control adds a new dimension to coherent control with applications in e.g nonlinear spectroscopy, microscopy and high-precision frequency metrology.**

In traditional coherent control experiments pulse shaping techniques (12) are used to steer light-matter interaction by manipulating the relative phases between different quantum paths leading to the same final state. Numerous control schemes have been shown in the past to provide frequency control and selectivity with a resolution exceeding the bandwidth of the individual pulses by 2-3 orders of magnitude. The vast majority of these schemes rely on a two-photon interaction with a single broadband shaped pulse. Even though the laser systems used in these experiments produce trains of pulses, decoherence effects due to the atomic motion wash out the interference between different pulses, and therefore blur the underlying atomic structure. Spatially, the interaction with a single shaped pulse creates a signal that is practically identical along the whole beam path.

Here we demonstrate how the control level of light-matter interaction is significantly enhanced by the interaction with multiple pulses. First, we show that the addition of a counter-propagating pulse changes the quantum interference and can even invert the properties of the excitation. In combination with pulse shaping techniques this geometry provides full spatial control and is able to produce complex excitation patterns over an extended region. Second, excitation from counter-propagating beams inherently reduces decoherence effects, allowing coherent accumulation of quantum interference over long pulse trains, even without the use of cooling techniques (13). Combined with the excellent phase stability of an optical frequency comb, this enables high-resolution excitation which is able to resolve the smallest features of the atomic structure.

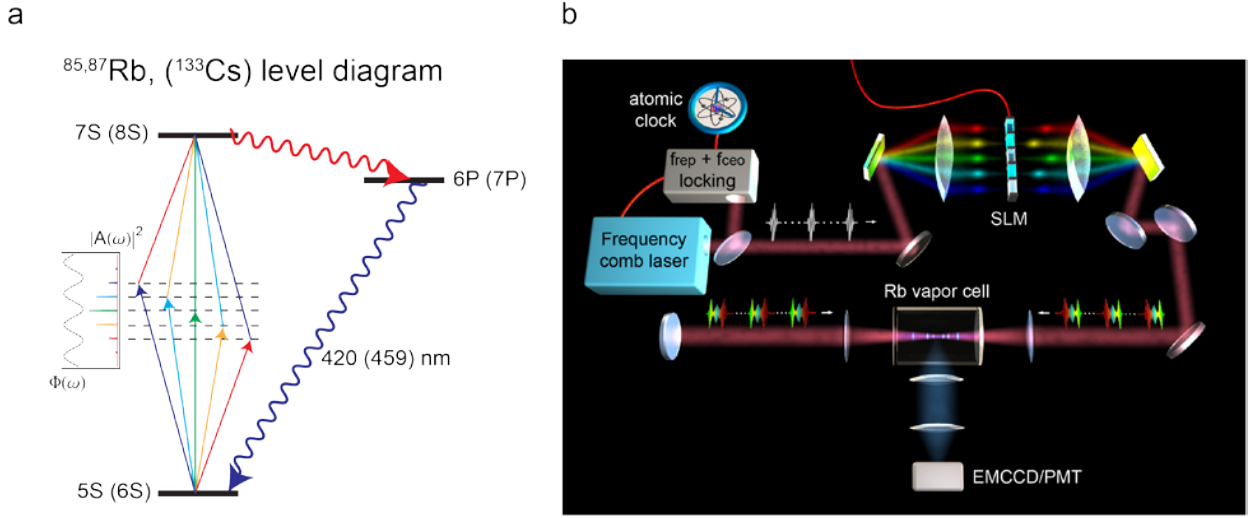

**Figure 1: Level diagram and a schematic of the experimental setup.** (**a**) Energy level diagram of atomic rubidium (cesium). Pairs of frequency comb modes excite the 5S-7S (6S-8S) transition. Fluorescence from the 6P (7P) state at 420 (459) nm is used for monitoring the excited state population. (**b**) Schematic of the experimental setup. Each pulse from a frequency comb laser is shaped using a spatial light modulator (SLM) in the Fourier plane of a zero dispersion 4f configuration. Every two consecutive pulses overlap in the middle of the vapour cel. The fluorescence pattern is imaged onto an EMCCD camera or a photomultiplier tube (PMT).

In order to gain insight in the properties of spatial control, we derive an analytical expression for the spatial excitation pattern. Consider the two-photon interaction of a simple two-level atom with a light field E(t), where the two photons originate from counter-propagating ultrashort pulses. Atoms at different positions experience the two counter-propagating pulses at different times with a delay Δt=z/c (z=0 is the point where the unshaped pulses overlap). The excitation probability from counter-propagating pulses can be written in the following way (complete derivation in the Supplementary Information):

$$S_{2p}(z) = \int_{-\infty}^{\infty} d\omega' \int_{-\infty}^{\infty} d\omega'' A(\omega')A(-\omega')A(\omega'')A(-\omega'') e^{i(\Phi(\omega')+\Phi(-\omega')-\Phi(\omega'')-\Phi(-\omega''))} e^{i\frac{2z}{c}(\omega'-\omega'')} \quad (1)$$

where A(ω) is the spectral amplitude and Φ(ω) is the spectral phase at a frequency ω of the laser field, relative to half the transition frequency ($\omega_0/2$). The appearance of the fields A(ω) and A(-ω) in the integral reflects the fact that many pairs of different spectral modes centered around $\omega_0/2$ are involved in the two-photon transition.

Some distinct properties of the excitation profile can be derived even without full evaluation of this integral. In the case of an anti-symmetric phase function Φ(ω)= −Φ(−ω), the phase dependent terms in the exponent of each integral cancel. The excitation probability will therefore be identical to the case of transform-limited pulses which is a similar behavior as observed in the single beam configuration (14). A new level of control is made possible by applying symmetric spectral phase masks, which results in the generation of complex spatial patterns. Remarkably, when Eq. 1 is integrated over the spatial dimension (simulating a position-insensitive detector), all phase terms drop out, resulting in optimal constructive interference. The spatial excitation pattern can thus be manipulated without affecting the total signal strength.

As a proof-of-principle we conducted measurements on various two-photon transitions in atomic rubidium and cesium (level diagrams are presented in Fig. 1a). The experimental setup is shown in Fig. 1b. Briefly, femtosecond pulses from a frequency comb laser are spectrally shaped in a standard phase-only 4f configuration. The shaped pulses are focused in a vapor cell containing either pure Rb or a Rb-Cs mixture. After passing the cell, the pulses are reflected back at a distance that matches the repetition frequency of the laser, so that every two consecutive pulses overlap in the middle of the vapor cell. Control over the atomic excitation is achieved by applying a spectral phase mask to the pulse shaper and tuning the frequency comb parameters $f_{rep}$ (repetition frequency) and $f_0$ (carrier-envelope offset frequency) which determine the absolute frequencies of the frequency comb modes through the comb equation $f_n = f_0 + n \times f_{rep}$ (10,11). The excited state population is monitored by detecting the 420 nm (459 nm) fluorescence upon spontaneous decay via the 6P (7P) state in rubidium (cesium).

First we applied a harmonic spectral phase mask, $\Phi(\omega) = \alpha \cos(\beta\omega + \phi)$, where $\alpha$ is the modulation depth, $\beta$ is the modulation frequency and $\phi$ is the modulation phase. We investigated the influence of this class of spectral phase masks on the excitation from single-sided and counter-propagating pulses. A pure Rb vapor cell was used, and the comb parameters were set to excite the $^{85}$Rb 5S(F=3)→ 7S(F=3) transition. As a reference a fluorescence image of the excitation pattern resulting from *transform-limited* pulses ($\alpha$=0) is shown in Fig. 2a. Transform-limited pulses lead to optimal constructive interference of single-sided excitation, giving rise to a strong spatially-independent signal. The localized signal in the center results from the spatial overlap of two counter-propagating transform-limited pulses.

A harmonic spectral phase mask alters the temporal intensity distribution of the pulses, breaking each pulse into a train of sub-pulses. In Fig. 2b a sine-modulated spectral phase ($\phi = \pi/2$, anti-symmetric relative to $\omega_0/2$) is applied with modulation depth of $\alpha$=1.2. An anti-symmetric spectral phase does not influence the single-sided excitation (14), which appears in our case as an unchanged, spatially independent background. We observe that the localised signal resulting from counter-propagating pulses is also unaffected by an anti-symmetric phase, as predicted from Eq.1. Although the sub-pulses overlap in multiple regions in space, the quantum interference is only constructive in the center.

Figure 2c shows an image of the excitation pattern when applying a symmetric phase mask ($\phi$=0) with the same modulation depth and modulation frequency as in Fig. 2b. Even though the temporal intensity distributions of the interacting pulses for Figs. 2b and 2c are identical, the resulting spatial excitation patterns differ substantially. For this symmetric phase mask, the single-sided excitation is completely eliminated, which is why such pulses have been named "dark pulses" (14). However, Fig. 2c shows that counter-propagating dark pulses actually give rise to a spatially structured two-photon excitation pattern where the sub-pulses overlap. We find that these counter-propagating dark pulses are not only not dark, but induce the same total excitation as transform-limited pulses.

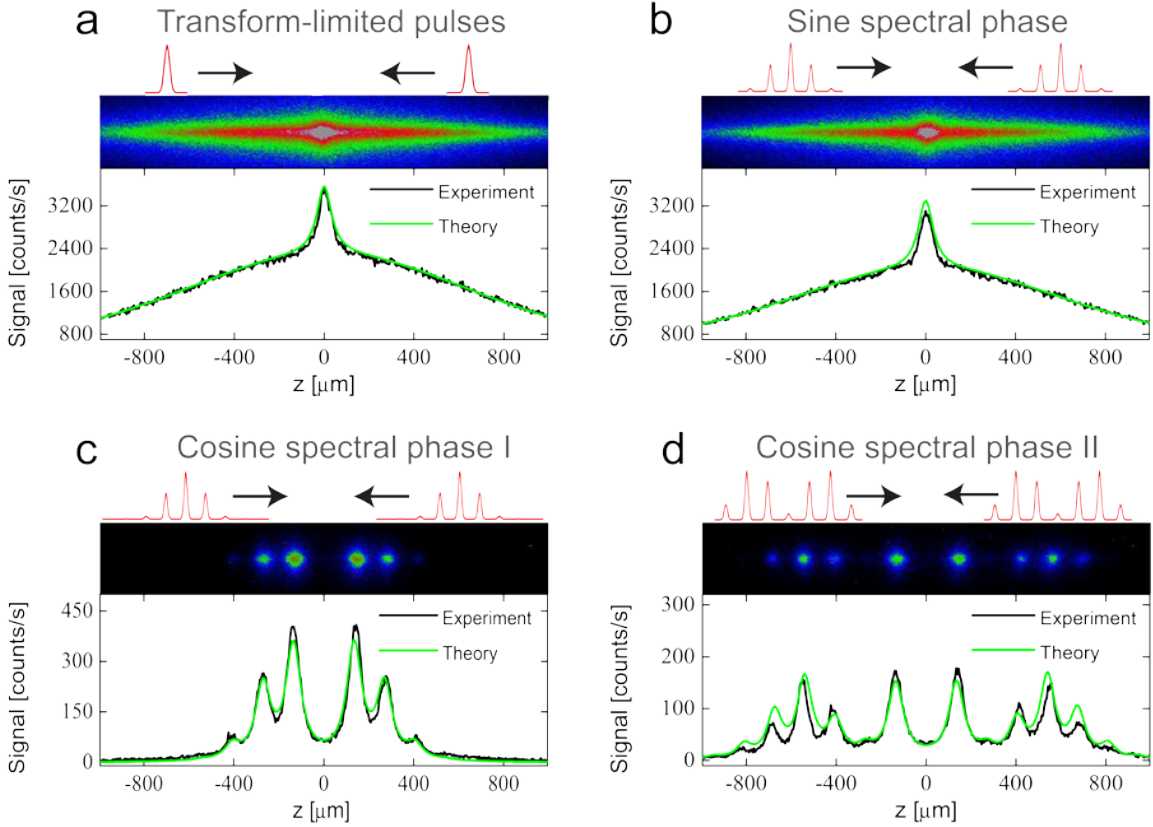

**Figure 2: Experimental demonstration of spatial coherent control, using four different phase masks.** (a) transform-limited pulses, (b) antisymmetric, sine modulated spectral phase ($\alpha =1.2$, $\beta = 900$ fs, $\varphi = \pi/2$), (c) symmetric, cosine modulated spectral phase ($\alpha =1.2$, $\beta = 900$ fs, $\varphi = 0$) and (d) symmetric, cosine modulated spectral phase with a larger amplitude ($\alpha =2.76$, $\beta = 900$ fs, $\varphi = 0$). For each figure we present the temporal intensity profile (depicted in red) of the counter-propagating pulses and the observed spatial excitation pattern. A cross-section of each fluorescence image is compared to a numerical evaluation of Eq. 1. The broadening of the data due to the imaging resolution and the atomic motion is taken into account in the theory by convolving all simulation curves with the same 80 µm FWHM Lorentzian distribution. A single scaling factor has been applied to all theoretical curves. For parts a and b the background from single-sided excitation is fitted to a Gaussian profile.

Variation of the modulation depth can be used as a "switch" to turn the single-sided signal on and off, but also to control the relative brightness of the individual excitation regions. In Fig. 2d the modulation depth was set to $\alpha=2.76$, where the single-sided signal is again completely eliminated. The fluorescence image displays additional excitation regions with different ratios between their relative intensities. In addition, the positions of the excitation regions can be controlled by varying the modulation frequency $\beta$, as can be seen in the Supplementary Figure 1.

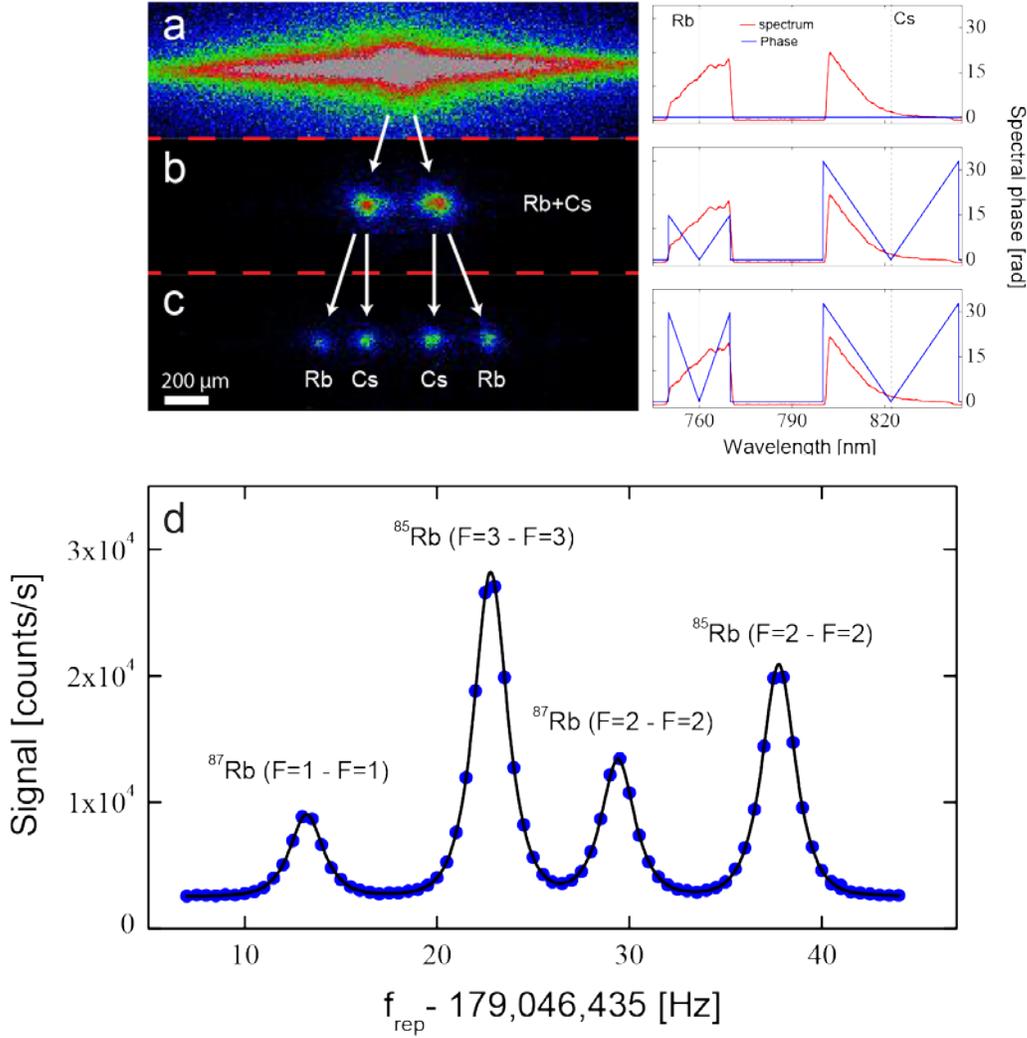

**Figure 3: Spatial coherent control of different atomic species.** Spatial excitation patterns using (**a**) transform-limited pulses. (**b**) Double V-shaped spectral phase mask ($\Phi(\omega)=\alpha|\omega-\omega_0/2|$) with $\alpha=0.5$ ps for both spectral regions. (**c**) Double V-shaped spectral phase mask with $\alpha=0.5$ ps for Cs and $\alpha=1.1$ ps for Rb. (**d**) A scan of the frequency comb over the 4 allowed rubidium transitions, where the solid line is a fit through the data. Spatial coherent control enables complete elimination of the Doppler-broadened background, significantly improving the signal to noise ratio, as well as selective excitation of a single atomic species.

In the second set of measurements we illustrate the versatility of spatial coherent control by spatially separating the excitation of different atomic species. We use a Rb-Cs mixture vapor cell to simultaneously excite the $^{85}$Rb 5S(F=3)→7S(F=3) transition in rubidium at 2×760 nm and the $^{133}$Cs 6S(F=4)→ 8S(F=4) transition in cesium at 2×822 nm. Figure 3a shows the excitation profile with transform-limited pulses. The signal is dominated by single-sided excitation where no distinction between the atomic species can be made. Subsequently, we employ a V-shaped spectral phase mask ($\Phi(\omega)=\alpha|\omega-\omega_0/2|$) that effectively eliminates the single-sided signal. By applying such a spectral phase mask to both spectral regions (centered around $\omega_{Rb}$=760 nm and $\omega_{Cs}$=822 nm) we create a background-free excitation pattern as seen in Fig. 3b. By changing the slope of the V-shaped phase masks, the excitation position of the individual species can be moved around independently. This results in controlled spatially separated excitation of rubidium and cesium atoms, as seen in Fig. 3c. We verify that the excitation within the various regions in Fig. 3c originates from different atomic species by scanning the frequency comb modes over the different atomic transitions. Such a scan reveals the hyperfine atomic structure, which is a unique fingerprint of the different atoms. The results are presented in Supplementary Figure 2, which clearly confirms the separation of atomic excitation.

The combination of background-free signal with the reduction of Doppler-broadening is essential for performing high-precision direct frequency comb spectroscopy (15-18). In Fig. 3d a scan of the 4 allowed 5S-7S transitions in rubidium is presented, which is recorded while scanning the comb repetition frequency. The excellent signal to noise ratio enables us to determine the hyperfine A constants of the excited state in $^{85}$Rb and $^{87}$Rb to be 94,686(7) kHz and 319,713(30) kHz, respectively. Simultaneously measuring both isotopes also permits accurate determination of the isotope shift. We find the isotope shift of the 5S → 7S transition to be 131,564(20) kHz. This presents the most accurate measurement of these values, outperforming other direct frequency comb measurements (19) and even CW laser measurements (20) (A detailed comparison is given in the Supplementary Table 1).

The combination of spatial and spectral coherent control opens new possibilities for a wide range of applications including nonlinear spectroscopy, microscopy and high-precision frequency metrology. For example, selective excitation of specific ions in a trap can be achieved with high precision. The ability of spatial coherent control to excite different transitions at distinct positions in space can for instance be exploited for spatially-dependent chemical bond formation or dissociation. Additionally, shaping only one of the two counter-propagating pulses can be used as a type of pump-probe measurement where the delay between the pump and the probe is mapped to the spatial dimension. Combined with spatially sensitive detection, signal from all possible delays can be acquired in a single shot. The combination of line by line shaping (21,22) with our principle of spatial coherent control could lead to optical arbitrary waveform generation in space and time.

**Methods:**

The frequency comb is based on a Ti:sapphire oscillator producing 20 fs pulses at 179 MHz repetition frequency. For the various measurements presented here we centered the spectrum to either 760 or 790 nm. Both comb parameters ($f_0$ and $f_{rep}$) were locked to a Global-Positioning System-disciplined Rb atomic clock. The data acquisition of the spectroscopy measurement was achieved by scanning the repetition frequency in steps of 0.5 Hz.

The shaper system is based on a zero-dispersion 2f-2f configuration, with two 1200 l/mm gratings and two 100 cm radius of curvature spherical mirrors. A 640-pixel computer-controlled SLM was placed at the Fourier plane to apply a phase-only spectral mask.

The excitation setup consists of a vapor cell containing either pure rubidium or a rubidium cesium mixture. The cell is heated to a temperature of 60 to 70 degrees Celsius in order to increase the pressure in the cell for increased signal strength. The laser beams are focused in the middle of the cell. For the spectroscopy measurements the focus size was chosen to be larger than 50 µm to avoid transit-time broadening.

**Acknowledgments:** S.W. acknowledges support from the Netherlands Organization for Scientific Research (NWO Veni grant 680-47-402). K.S.E.E. acknowledges support from NWO through a VICI grant (680-47-310), the Foundation for Fundamental Research on Matter (FOM) through its program "Broken Mirrors and Drifting Constants", and Laserlab Europe (JRA ALADIN and INREX). The authors thank Misha Sheinman for helpful discussions in the development of the theoretical model.

**Author contributions:**

K.S.E.E. conceived the V-shape phase concept for frequency comb spectroscopy. I.B extended this concept to full spatial coherent control and provided the theoretical description. I.B. performed the experiments (with assistance from S.W and K.S.E.E.) and the data analysis. K.S.E.E. supervised the project. All authors participated in the design of the experiments, interpretation of the results and writing of the manuscript.

**Additional information:**

The authors declare no competing financial interests.

# Supplementary information:

## Derivation of Eq. 1

The excitation probability of a two-photon transition from a single pulse was derived in (1):

$$S_{1p} = \left| \int_{-\infty}^{\infty} d\omega A(\omega) A(-\omega) e^{i(\Phi(\omega)+\Phi(-\omega))} \right|^2 \tag{S1}$$

In the present work we are interested in the excitation due to counter-propagating pulses. Atomic excitation occurs due to the perturbation of the atomic wave function by an electromagnetic field. The direction of the electromagnetic field does not affect the excitation probability (ignoring effects of momentum transfer). Therefore, at each point z the excitation probability is the same as for co-propagating pulses with a time delay $\Delta t = \frac{2z}{c}$. This time delay is incorporated by adding a linear spectral phase $\Phi(\omega) \to \Phi(\omega) \pm z\omega/c$. Equation S1 then becomes:

$$S_{2p}(z) = \left| \int_{-\infty}^{\infty} d\omega A(\omega) A(-\omega) e^{i(\Phi(\omega)+\Phi(-\omega)+2\frac{z\omega}{c})} \right|^2 \tag{S2}$$

We rewrite Eq. S2 as a product of the integral with its complex conjugate and rearrange the integrals to acquire the following result:

$$S_{2p}(z) = \int_{-\infty}^{\infty} d\omega' \int_{-\infty}^{\infty} d\omega'' A(\omega')A(-\omega')A(\omega'')A(-\omega'') e^{i(\Phi(\omega')+\Phi(-\omega')-\Phi(\omega'')-\Phi(-\omega''))} e^{i\frac{2z}{c}(\omega'-\omega'')} \tag{S3}$$

If this signal is observed with a position-insensitive detector, the total signal is found by integration of Eq. S3 over the spatial coordinate *z*. However, the last term in Eq. S3 represents the definition of a delta distribution when integrating over *z*. Therefore only components where $\omega' = \omega''$ lead to a nonzero contribution in this case. The total signal is then equal to:

$$S_{total} = \int_{-\infty}^{\infty} dz S_{2p}(z) = \int_{-\infty}^{\infty} d\omega A(\omega)^2 A(-\omega)^2 \tag{S4}$$

The consequence of this result is that the total (spatially integrated) signal due to counter-propagating pulses is independent of the specific form of the applied spectral phase.

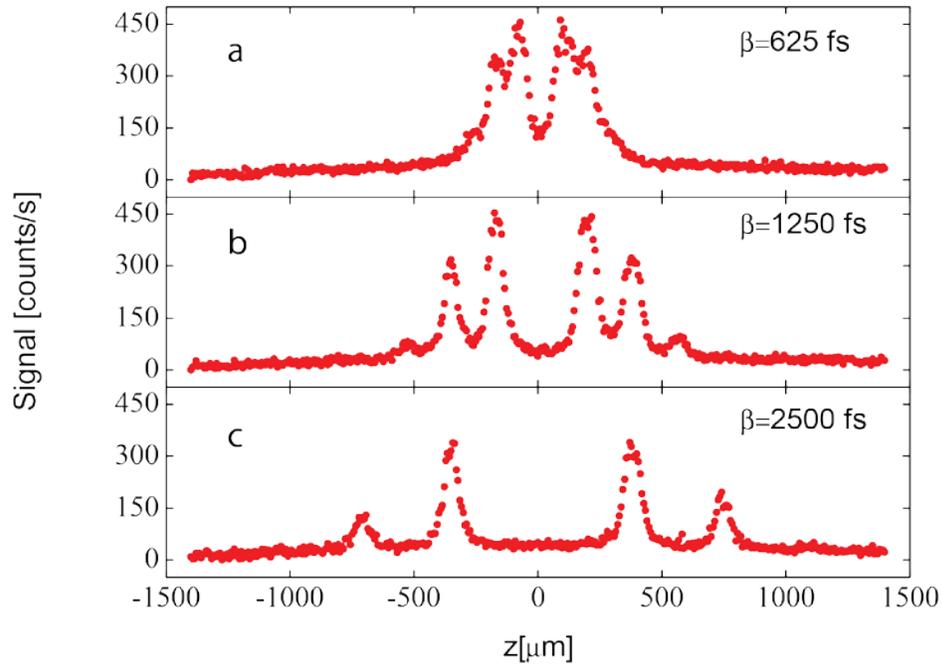

**Supplementary Figure 1: Influence of the modulation frequency on the spatial excitation patterns.** The spectral phase mask applied to the pulses is of the form $\Phi(\omega)=1.2\cos(\beta\omega)$. Such a phase mask splits each pulse into a train of sub-pulses, where the modulation frequency $\beta$ determines the time delay between the sub-pulses. The modulation amplitude was set to 1.2 in order to eliminate the background from single-sided excitation. The modulation frequency is (a) 625 fs, (b) 1250 fs and (c) 2500 fs. All graphs were acquired by taking a cross-section of fluorescence images, in the same way as in Fig. 2.

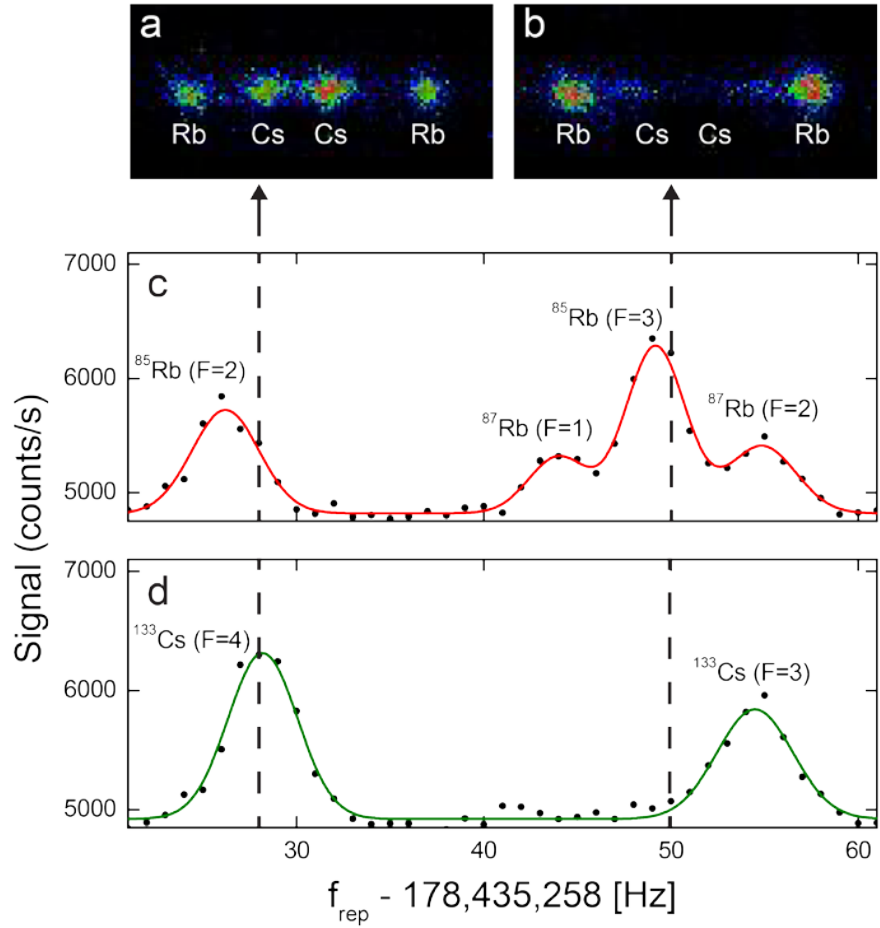

**Supplementary Figure 2: High-precision spatial and spectral control of multiple atomic species.**
(a,b) Spatial excitation patterns of specific hyperfine transitions in atomic rubidium and cesium. The two images are taken with two different sets of comb parameters as indicated by the dashed lines. (c,d) frequency scan over the atomic transitions in rubidium and cesium, respectively. These results show that the excitation of the two atomic species is spatially separated. The signal is obtained directly from the EMCCD camera, which leads to a relatively high background due to noise resulting from the high EMCCD gain setting used in this experiment.

|  | Marian *et al.* (2) | Chui *et al.* (3) | Present work |
|---|---|---|---|
| $^{85}$Rb *A* constant | - | 94,658 (19) | 94,686 (7) |
| $^{87}$Rb *A* constant | 319,702 (65) | 319,759 (28) | 319,713 (30) |
| $^{87}$Rb- $^{85}$Rb Isotope shift | - | 131,567 (73) | 131,564 (20) |

**Supplementary Table 1: Comparison of this work with previous measurements (all values in kHz).** The various measurements schemes are: direct frequency comb resonantly-enhanced excitation by Marian *et al.* (2), CW laser excitation with equal photons by Chui *et al.* (3), and direct frequency comb excitation without resonant excitation (present work).